\documentclass[conference]{IEEEtran}
\usepackage[dvipdfm]{graphicx}
\usepackage[T1]{fontenc}
\usepackage[applemac]{inputenx}     
\usepackage{cite}
\usepackage[usenames,dvipsnames]{color}
\usepackage{amsmath}
\usepackage{color}
\usepackage{epstopdf}%
\usepackage{amsfonts}%
\usepackage{subfigure}
\usepackage{amssymb,bold-extra}%
\usepackage{graphicx}%
\usepackage{multicol}%
\usepackage{verbatim}%
\usepackage{enumerate}
\usepackage{amsthm}
\usepackage{eucal}

\begin{document}

\title{Sparse Reconstruction-Based Detection of Spatial Dimension Holes in Cognitive Radio Networks}

\author{\large Yahya H. Ezzeldin$^*$, Radwa A. Sultan$^*$, and Karim G. Seddik$^\ddag$\\ [.1in]
\normalsize  \begin{tabular}{c}
$^*$Electrical Engineering Department, Alexandria University, Alexandria, Egypt. \\
$^\ddag$Electronics Engineering Department, American University in Cairo, AUC Avenue, New Cairo, Egypt.\\
 Email: yahya.ezzeldin@ieee.org, radwasultan@ieee.org, kseddik@aucegypt.edu
\end{tabular}
 }
 \maketitle
\begin{abstract}
In this paper, we investigate a spectrum-sensing algorithm for detecting spatial dimension holes in Multiple-Input Multiple-Output (MIMO) transmissions for OFDM systems using Compressive Sensing (CS) tools. This extends the energy detector to allow for detecting transmission opportunities even if the band is already energy filled. 
We show that the task described above is not performed efficiently by regular MIMO decoders (such as MMSE decoder) due to possible sparsity in the transmit signal. Since CS reconstruction tools take into account the sparsity order of the signal, they are more efficient in detecting the activity of the users. Building on successful activity detection by the CS detector, we show that the use of a CS-aided MMSE decoder yields better performance rather than using either CS-based or MMSE decoders separately. 	
\end{abstract}
\section{Introduction}\label{Int}
Recent statistical measures by the Federal Communications Commission (FCC) showed the fixed assigned bands are highly underutilized \cite{FCC}. Cognitive radios (CR) \cite{haykin2005cognitive} appeared as a solution to the great inefficiency in bandwidth utilization. To overcome this, the CR nodes are required to have spectrum-sensing functions, and harbour dynamic and agile spectrum access functions that allow it to tap in on sensing idle activity and tap out of spectrum band on sensing return of activity. 

The overlay secondary users (SUs) in cognitive networks have an incentive to sense primary activity before accessing any band. SUs need to make sure that there is no PU occupying the band. If occupied, SUs cannot transmit their signals in this band. An exception for the previous, however holds, if the SU transmission can occupy un-tapped dimensions in this band or propagate along directions un-effective to PU receiver that show no effect to the PU decoded data. To match this incentive, spectrum-sensing techniques have been developed, namely, energy detector, matched-filter detector and feature detector \cite{akyildiz2006next}. Of the three detectors, energy detector has the benefit of being thoroughly generic. 

Energy detector requires no information about the signal form or the modulation technique used. It is unfortunate however, that due to this blind activity detection, energy detectors can mark some bands as occupied (or busy) while there still exists some opportunities for SU transmission in this band. One such possible opportunity is making use of the fact that primary receiver antennas need to be greater than or equal to the primary transmit antennas for correct decoding. In case receiver antennas are greater, the spatial dimensions are not fully utilized (i.e., more transmitters can be supported with the information still being decodable). Since the opportunity in this case is in the spatial dimension, we name the occurrence of such event as \textit{``spatial dimension holes''}. {\color[rgb]{0,0,0}{In order to utilize these \textit{spatial dimension holes}, the secondary user needs to be aware of the number of receiver antennas (which can be provided by the PU as a metric ) versus the number of active transmitters. Knowing the number of active transmitters (without regular information from the primary system) is}} a challenging problem for conventional systems that rely on matrix inversion techniques such as Zero-Forcing and MMSE Detectors. 

Recently, Compressive Sensing (CS)~\cite{Candes,Eldar,l2l1,donoho}  has been adopted by the signal processing community as a means for detecting sparsity patterns and recovering sparse signals. The use of CS for cognitive networks is not new and has been proposed in~\cite{cs-lit1,cs-lit3,cs-lit4}. However, all these publications consider using CS for sensing activity in wideband channels. To the best of our knowledge, using CS tools for exploiting transmission opportunities in spatial domain with limited antenna resources has not been tackled previously and this is the main incentive behind this paper.

In this paper, the uplink of a multiuser MIMO (MU-MIMO) system is considered\footnote{MU-MIMO is one of the transmission modes defined in the LTE standard.}. The primary base station (BS) is equipped with multiple antennas which enables the simultaneous transmission from multiple primary users. The primary BS can demodulate a number of primary users uplink streams, sharing the same frequency resources, that is less than or equal to the number of antennas at the BS. If the number of primary uplink streams is less than the number of antennas at the primary BS this will provide a ``spatial'' hole that can be occupied by the secondary users. Having more antennas at the primary BS allows for the transmission by the secondary users in the spatial holes and the interference can be separated at the primary BS. 

In this paper, we consider the use of the CS tools to detect the number of active primary users and hence detect the ``spatial'' spectrum holes that can be accessed by the secondary users. We also consider the use of the CS tools as well as the MMSE MIMO detector at the secondary network for decoding the primary users data for possible relaying of the primary users data.


\section{System Model}\label{sysmod}
\textit{Notations}:
Throughout the paper we refer to vectors with bold lower cases such as $\mathbf{x}$. Matrices are referred to with bold upper cases such as $\mathbf{A}_{M \times N}$, where $\mathbf{A}$ is a matrix of size $M\times N$. $\mathbf{diag}(\mathbf{y})$ refers to the diagonal matrix whose diagonal elements are the elements of the vector $\mathbf{y}$. Due to size limitations, we will sometimes refer to $\mathbf{diag}(\mathbf{y})$ as $\mathbf{D}(\mathbf{y})$. Hermitian of a matrix $\mathbf{A}$ is denoted by $\mathbf{A}^\mathcal{H}$. $\mathbf{A}^T$ denotes the transpose of the matrix $\mathbf{A}$. We use $\mathbb{R}^N$ to denote the vector space of the $N\times 1$ real vectors; $\mathbb{C}^N$ is defined similarly to be the vector space of the $N\times 1$ complex vectors.
\subsection{Primary System Model}\label{Primary_sysmod}
We consider a single cell OFDM primary system with $N_P$ users collaboratively communicating with a single primary base station receiver (PR-BS) in a multiuser MIMO (MU-MIMO) setting. The $N_P$ transmitters spatially share the spectrum by employing a virtual MIMO setting. The number of receiving antennas at PR-BS $N_{BS}$ should be greater than or equal to $N_P$, where $N_P$ is the maximum number of active primary users simultaneously assigned to the same subcarriers. 


\begin{figure}
	\centering
	\includegraphics[scale=0.28]{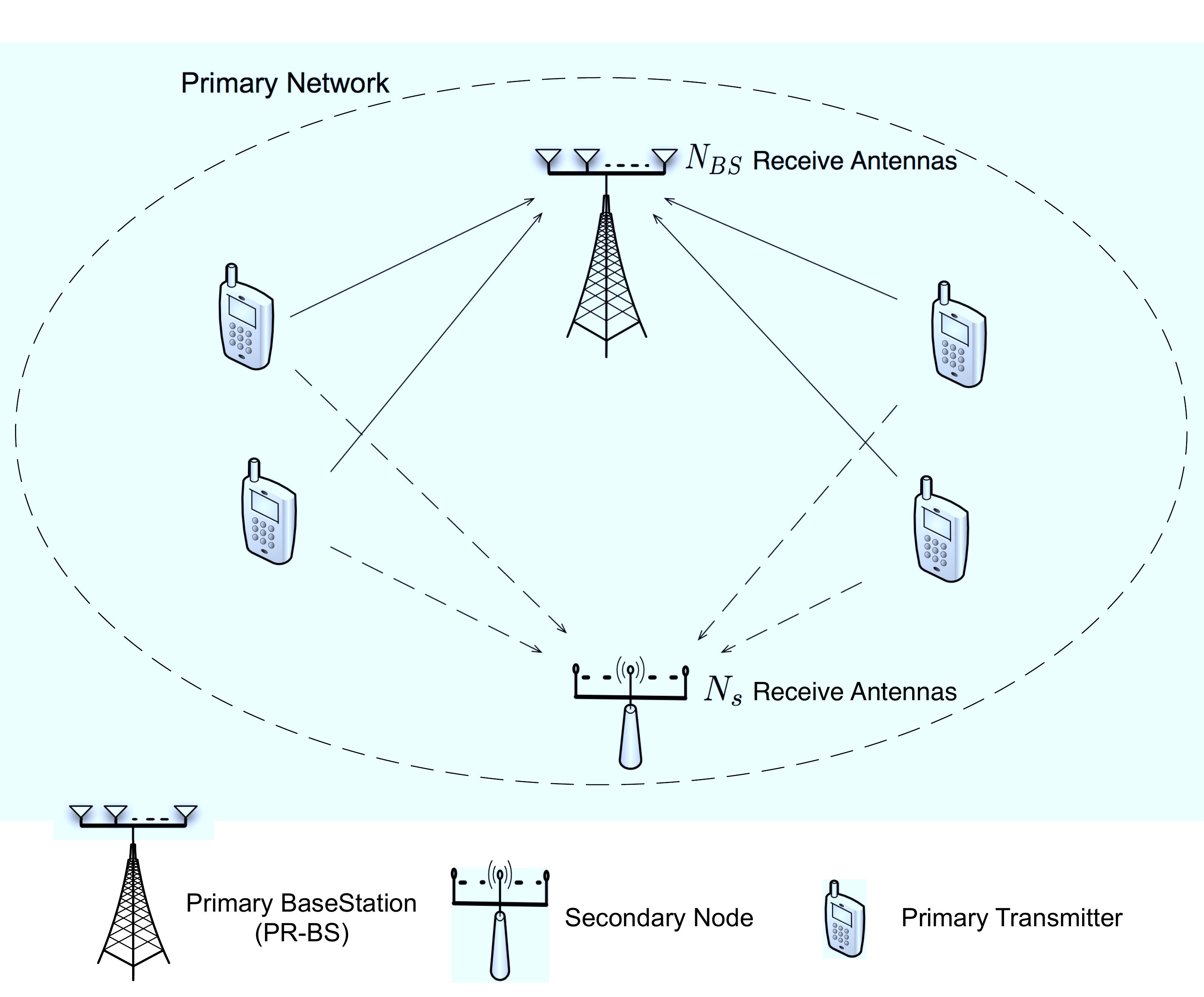}
	\caption{System Model.}
	\label{sysmodel}
\end{figure}
Let $\mathbf{x}_i \in \mathbb{C}^{L}$ be a transmit OFDM symbol from the primary user $i$. The elements in $\mathbf{x}_i$ are fed from IQ lattice constellations such as: QPSK, 16QAM or 64QAM. Moreover, $\mathbf{x}_i$ satisfies an average transmit power constraint of {\color[rgb]{0,0,0}{unity per subcarrier. Therefore the total power constraint for the vector is}} $E\lbrace \Vert \mathbf{x}_i \Vert_{\ell_2} \rbrace \leq L${\color[rgb]{0,0,0}{, where $L$ is the number of subcarriers used for transmission}}.

\subsection{Secondary System Model}\label{Sec_sysmod}
We consider an OFDM secondary user (SU) with $N_S$ antennas. The secondary user uses its $N_S$ peripherals to sense the degree of spectrum usage in the spatial domain. For the model presented in this paper, we consider two different scenarios that the SU can follow. 
\begin{enumerate}
 \item The SU attempts to make use of the free transmission dimension to transmit its own information simultaneously with the active primary users. 
 \item The SU detects the active primary users, decodes their transmitted symbols for possible relaying of the PU data.
 \end{enumerate}
It is assumed that the channel between the $N_P$ primary users and the $N_S$ secondary user antennas is perfectly known at the SU. This assumption is not far from practical because channel estimation can be performed by the SU using the reference signals (RS) transmission by the primary users to the PR-BS. The primary system performs channel estimation over a number of transmission slots. Assuming slowly fading channel, the estimates for the channel between the primary and the secondary system are assumed to be valid over a number of transmission slots.

The channel coefficients $\mathbf{h}^s_{j,i}$ from the $i$-th primary user to the $j$-th {\color[rgb]{0,0,0}{antenna on the}} secondary user are modelled in time domain as a multipath fading channel with independent taps and each tap is modelled as Rayleigh fading where the sum of the taps' variances equals 1. The received signal on the $k$-th subcarrier at the $j$-th {\color[rgb]{0,0,0}{antenna}} is
\begin{equation}
\mathbf{y}_j(k) = \sum_{i=1}^{N_P} \mathbf{h}^s_{j,i}(k) \cdot \mathbf{x}_i(k) + \mathbf{n}_j(k) \;\;,\;\;k = 1,\dots, L,
\end{equation}
where $\mathbf{y}_{j}$ is the ($L \times 1$) vector received at the $j$-th at the $j$-th {\color[rgb]{0,0,0}{antenna on the secondary user}} and $\mathbf{n}_j$ is a vector of i.i.d. complex Gaussian noise samples received at the $j$-th {\color[rgb]{0,0,0}{antenna}} which have zero mean and variance $\sigma_n^2, \mathcal{CN}\ (0, \sigma_n^2$). The SNR for the system is therefore, $1/\sigma_n^2$.
This can be rewritten to include all subcarriers as
\begin{equation}
\mathbf{y}_j = \sum_{i=1}^{N_P} \mathbf{D} (\mathbf{h}^s_{j,i})\; \mathbf{x}_i + \mathbf{n}_j .
\end{equation}
By combining the received vectors from all of the $N_S$ receivers, we get
\begin{equation}
\label{bigeqn}
\mathbf{y} = \underbrace{\left[ \begin{smallmatrix}
 \mathbf{D} (\mathbf{h}^s_{1,1}) & \mathbf{D} (\mathbf{h}^s_{1,2})& \cdots &\mathbf{D} (\mathbf{h}^s_{1,N_P}) \\
\mathbf{D} (\mathbf{h}^s_{2,1}) & \mathbf{D} (\mathbf{h}^s_{2,2}) & \cdots & \mathbf{D} (\mathbf{h}^s_{2,N_P}) \\
\vdots & \vdots & \vdots & \vdots\\
\mathbf{D} (\mathbf{h}^s_{N_S,1}) & \mathbf{D} (\mathbf{h}^s_{N_S,2}) & \cdots & \mathbf{D} (\mathbf{h}^s_{N_S,N_P}) \\
\end{smallmatrix} \right]}_{\mathbf{H}} \;
\left[ \begin{smallmatrix}
 \mathbf{x}_1\\
 \mathbf{x}_2\\
 \vdots \\
 \mathbf{x}_{N_P} \\
\end{smallmatrix} \right]+
\left[ \begin{smallmatrix}
 \mathbf{n}_1\\
 \mathbf{n}_2\\
 \vdots \\
 \mathbf{n}_{N_S} \\
\end{smallmatrix} \right],
\end{equation}
where the received signal at the secondary node is given by
\begin{equation}
\begin{split}
\nonumber
\mathbf{y} &= [\mathbf{y}_1^T\;\;\mathbf{y}_2^T\;\dots\;\; \mathbf{y}_{N_S}^T]^T
\end{split}
\end{equation}
and $\mathbf{H}$ is the channel matrix.

\section{Proposed Sensing Strategy}\label{strategy}
Before discussing the proposed sensing strategy, we refer to important compressive sensing results used in our work.

\subsection{Compressive Sensing}\label{csintro}
Compressive Sensing is a technique for reconstructing sparse vector $\mathbf{v} \in \mathbb{R}^N$ from a small set of compressive measurements. Signal $\mathbf{v}$ is denoted as $K$-sparse if at most $K$ elements of $\mathbf{v}$ are non-zeros.
Pioneered by Candes et al.~\cite{Candes}, it has been demonstrated that reconstruction of $\mathbf{v}$ in noisy conditions, ($\mathbf{r} = \mathbf{A}\mathbf{v} + \mathbf{n},\;\mathbf{r}\in \mathbb{R}^M$),  is unique with negligible probability of error by solving the $\ell_{1}$-minimization problem 
\begin{equation}
\begin{split}
P1 : \hspace*{5em} \min_{\tilde{\mathbf{v}}}\; \Vert \tilde{\mathbf{v}} \Vert_{\ell_1}  \hspace*{8em}
\\ \text{subject to}\;:\;\; \Vert \mathbf{r}-\mathbf{A}\tilde{\mathbf{v}} \Vert_{\ell_2} \leq \epsilon 
\end{split}
\end{equation}
where $\epsilon$ is a term that bounds the tolerable noise energy in the estimated signal $\tilde{\mathbf{v}}$. {\color[rgb]{0,0,0}{Although the $M \times N$ matrix $\mathbf{A}$ is rank deficient and loses information, it can be shown to preserve the information in sparse and compressible signals if it satisfies the so-called \textit{restricted isometry property} (RIP) \cite{Candes}. Checking whether a matrix satisfies the RIP condition is an NP-Complete problem \cite{natarajan1995sparse}, however, for random matrices whose entries are independent and identically distributed (i.i.d.) Gaussian, the RIP condition is satisfied with high probability given that $M \geq K\log(N/K)$ \cite{Candes,model_CS}.}} Currently, great research effort has been invested in improving the computational complexity of compressive sensing reconstruction techniques. Techniques such as ``subspace pursuit'' and ``orthogonal matching pursuit'' have been developed that exhibit computational complexity of $\mathcal{O}(NM)$ and $\mathcal{O}(N\log M)$, respectively. For the context of this paper, and due to space limitations, we do not consider any specific reconstruction technique, however, any of the formerly mentioned techniques can be used for reconstruction. We refer the reader to \cite{SP_CS,OMP_CS,SOMP_CS} for further discussion regarding the implementation and complexity analysis for these techniques.

\subsection{Block Sparse Reconstruction}\label{blocksparse-desc}
A further extension to the generic compressive sensing discussed previously is by making use of additional structure properties in the signal \cite{model_CS}. Some signals have non-zero elements arranged in the form of blocks and hence denoted $Block\;Sparse\;Signals$.
Block Sparse signals representation comes naturally in multi-channel signals which are in question in this paper. 

In \cite{Eldar,l2l1}, Eldar and Stojnic demonstrated that extending the $\ell_1$-minimization algorithm proposed in~\cite{Candes} by explicitly making use of block-sparsity yields better reconstruction properties than treating the signal as being conventionally just sparse. To describe a block sparse vector of length $N$, we will assume that integers $n$ and $d$ are chosen such that $n=N/d$ is an integer as well. In this context, $d$ represents the block size and $n$ is the number of blocks. A signal is $k$-block-sparse if at most $k=K/d$ blocks are non-zero.  The extended problem therefore becomes
\begin{equation}
\label{blocksparse1}
\begin{split}
P2 : \hspace*{4em} \min_{\tilde{\mathbf{v}}}\; \sum_{i=1}^n \Vert \tilde{\mathbf{v}}_{(i-1)d+1:id} \Vert_{\ell_2} \hspace*{2em}\;
\\ \text{subject to}\;:\;\; \Vert \mathbf{r}-\mathbf{A}\tilde{\mathbf{v}} \Vert_{\ell_2} \leq \epsilon 
\end{split}
\end{equation}
where $\tilde{\mathbf{v}}_{(i-1)d+1:id}$ represent the elements of vector $\tilde{\mathbf{v}}$ from indices $(i-1)d+1$ to $id$.

{\color[rgb]{0,0,0}{For the previously described model for the signal the relaxed condition for the number of measurements required to satisfy the RIP condition becomes $M \geq K + k\log(N/k)$ \cite{model_CS}. This poses significant improvement over the the $M \geq K\log(N/K)$ required with signals without block sparse structure. This translates to $M = \mathcal{O}(K)$ as the size of the block increases.}}
\subsection{Proposed Sensing Algorithm}\label{proposedsensing}
In the proposed sensing technique, we model the concatenated data vector from different primary users $\mathbf{x} = [\mathbf{x}_1^T\;\;\mathbf{x}_2^T\;\dots\;\mathbf{x}^T_{N_P}]^T$ as $K$-block-sparse where $K \leq N_P$. {\color[rgb]{0,0,0}{Since a transmitting user can either utilize all the subcarriers $L$ in a resource block or none at all, the signal can be broken down into blocks each of size $L \times 1$ and therefore modelled as a block-sparse signal}}. The reason for such modelling flows Let $\mathbb{A}$ be the modulation alphabet used. We define an extended modulation alphabet $\mathbb{A}'= \mathbb{A} \bigcup \lbrace 0 \rbrace$ to allow for the possible state of no-transmission in the constellation. We follow these steps to detect the activity and decode the symbols transmitted by the users as follows.

\subsubsection{\textbf{Activity Pattern Detection}}
To detect spatial activity, we customize the convex problem \eqref{blocksparse1}, with $\tilde{\mathbf{x}} \in \mathbb{C}^{N}$ being the target vector to be recovered. We then infer the state of the $i$-th user from the different entries in the vector $\tilde{\mathbf{x}}_i$. The algorithm is described in steps as follows.
\begin{enumerate}[\bfseries(Step 1)]
\item Solve the convex problem \eqref{blocksparse} for $\tilde{\mathbf{x}} \in \mathbb{C}^{N_P}$
\begin{equation}
\label{blocksparse}
\begin{split}
\min_{\tilde{\mathbf{x}}}\; \sum_{i=1}^{N_{P}} \Vert \tilde{\mathbf{x}}_i\Vert_{\ell_2} \hspace*{5em}\;
\\ \text{subject to}\;:\;\; \Vert \mathbf{y}-\mathbf{H}\tilde{\mathbf{x}} \Vert_{\ell_2} \leq \dfrac{1}{2}\sigma_n^2 \cdot N_S \cdot L
\end{split}
\end{equation}
\item {\color[rgb]{0,0,0}{Make a binary decision for each element in $\tilde{\textbf{x}}$ as follows: 
\begin{equation*}
\nonumber
\hat{\mathbf{x}}(i) = \left\{ 
  \begin{array}{r r}	
    \tilde{\mathbf{x}}(i) & \quad \text{if \;$\vert\tilde{\mathbf{x}}(i)\vert \geq \rho$}\\
    0 & \quad \text{otherwise,}\\
  \end{array} \right.
\end{equation*}
where $\hat{\mathbf{x}}$ denotes the output from the decision operation.}} The threshold $\rho$ is shown in Fig \ref{constellation} {\color[rgb]{0,0,0}{and it represents the plane of points that are equidistant form 0 and the nearest symbol in the alphabet $\mathbb{A}'$}}. In this step, the values in the indices defined by $P$ are set to zero and marked as energy empty. 
If the constellation used by the PU is known the points in $\tilde{\mathbf{x}}$ to be set to zero can be selected as the members of the set $P$ such that
\begin{equation}
\nonumber
P = \{\;i\;\vert\;\;\vert\tilde{\mathbf{x}}(i)\vert < \min_{x_c \in \mathbb{A}}\vert\tilde{\mathbf{x}}(i) - x_c\vert \}.
\end{equation}
This is similar to a minimum distance (MD) rule over the extended constellation (with the zero constellation point included). 

\begin{figure}
	\centering
	\includegraphics[scale=0.4]{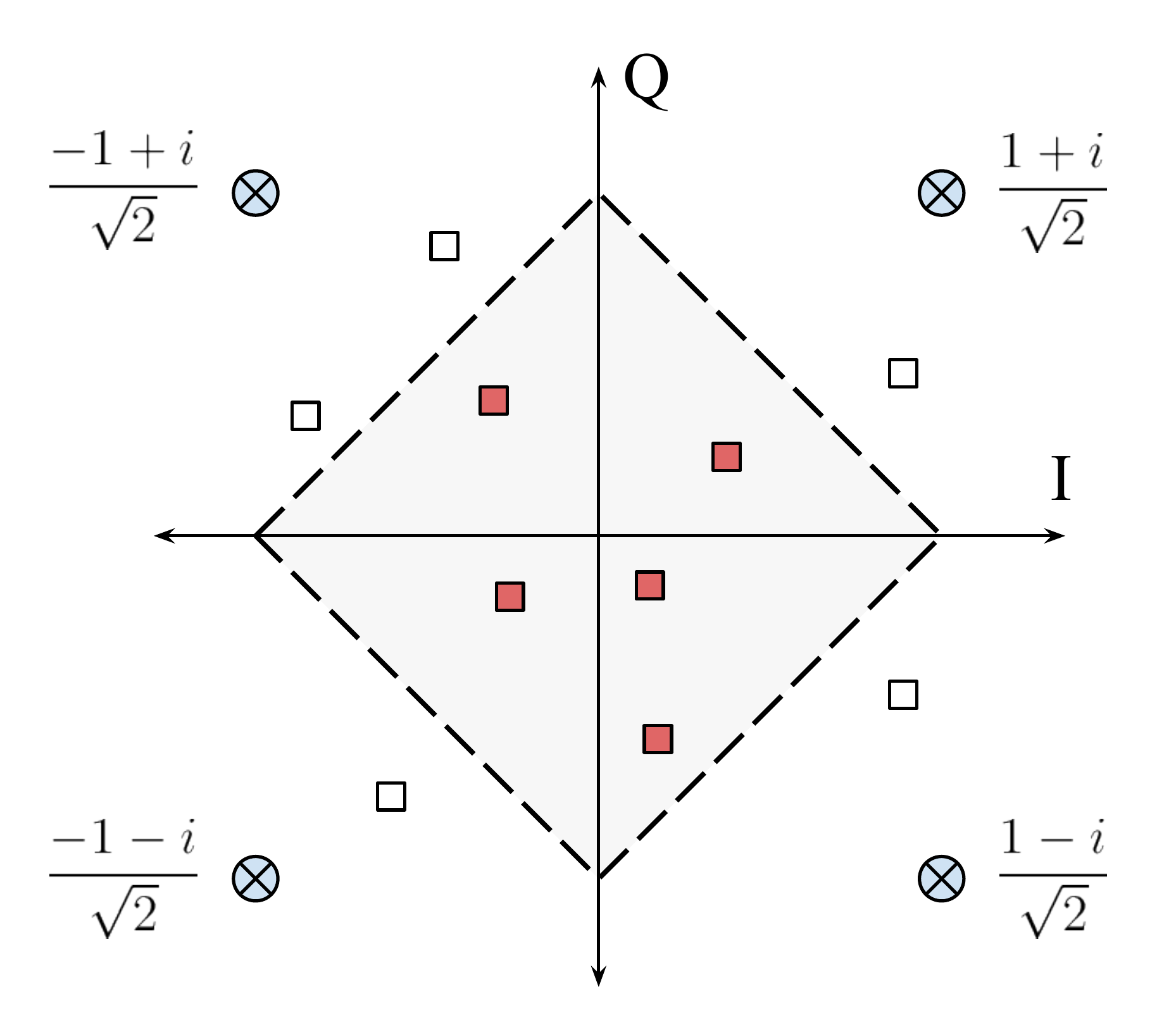}
	\caption{QPSK Constellation with threshold contour showing received signals points (squares) and the signals that are to be zeroed out in red color.}
	\label{constellation}
\end{figure}

\vspace{0.7em}
\item Construct the activity vector $\mathbf{a}$ where:
\begin{equation*}
\mathbf{a} = \left[a_1\;\;a_2\;\dots\;\;a_{N_p} \right]^T
\end{equation*}
\begin{equation}
\nonumber
a_i = \left\{ 
  \begin{array}{r r}	
    1 & \quad \text{if \;$\Vert \hat{\mathbf{x}}_i \Vert_{\ell_0}\geq L/2$}\\
    0 & \quad \text{otherwise,}\\
  \end{array} \right.
\end{equation}
where $\Vert \cdot \Vert_{\ell_0}$ denotes the $\ell_0$-norm. The vector $\mathbf{a}$ is characterized by sparsity pattern $(S)$ which define the indices of the non-zero elements of $\mathbf{a}$ \\
\end{enumerate}
\subsubsection{\textbf{Demodulating Active Users Symbols}}Once the activity pattern has been detected,  we use an MMSE Detector to detect the sparse subset of the transmitted signal, $\tilde{x}_s$:
\begin{equation}
\label{MMSEsparse}
\tilde{\mathbf{x}}_s = (\mathbf{H}^\mathcal{H}_s \mathbf{H}_s +\sigma_n^2\mathbf{I})^{-1}\mathbf{H}^\mathcal{H}_s \;.\; \mathbf{y}
\end{equation}
where, $\mathbf{H}_s$ is the subset matrix of $\mathbf{H}$ formed by the columns, indexed by the sparsity pattern $(S)$, of the active users. i.e.: $\mathbf{H}_s$ only contains the columns corresponding to the active users after discarding inactive columns from $\mathbf{H}$. Matrix $\mathbf{I}$ is the $|S| L \times |S| L$ identity matrix where $|S|$ is the cardinality of the set $(S)$. The MMSE equalized signal $\tilde{\mathbf{x}}_s$ is approximated to the nearest constellation point {\color[rgb]{0,0,0}{in $\mathbb{A}$}} using a minimum distance decision device.
\section{Simulation Results and Evaluation}\label{results}
In this section, we present the numerical results for the proposed spatial activity detector. We consider $N_p=8$ primary user transmitters  with $N_a$ active transmitters communicating with a single PR-BS with $N_{BS}=8$ receiving antennas. The secondary user attempts to detect the activity using $N_S$ receivers. We run simulations for 2 active users out of $N_P = 8$ and number of receiving antennas at the SU is $N_S = 4,6,8$. The number of subcarriers that are assigned simultaneously to the $8$ users is 72 subcarriers and the transmitted symbols are QPSK modulated. The channel between  the PU and SU are all modelled as a 10-tap channel where the taps are of equal variance (the variance of each tap is $1/10$).

\begin{figure}
	\centering
	\includegraphics[width=0.46\textwidth]{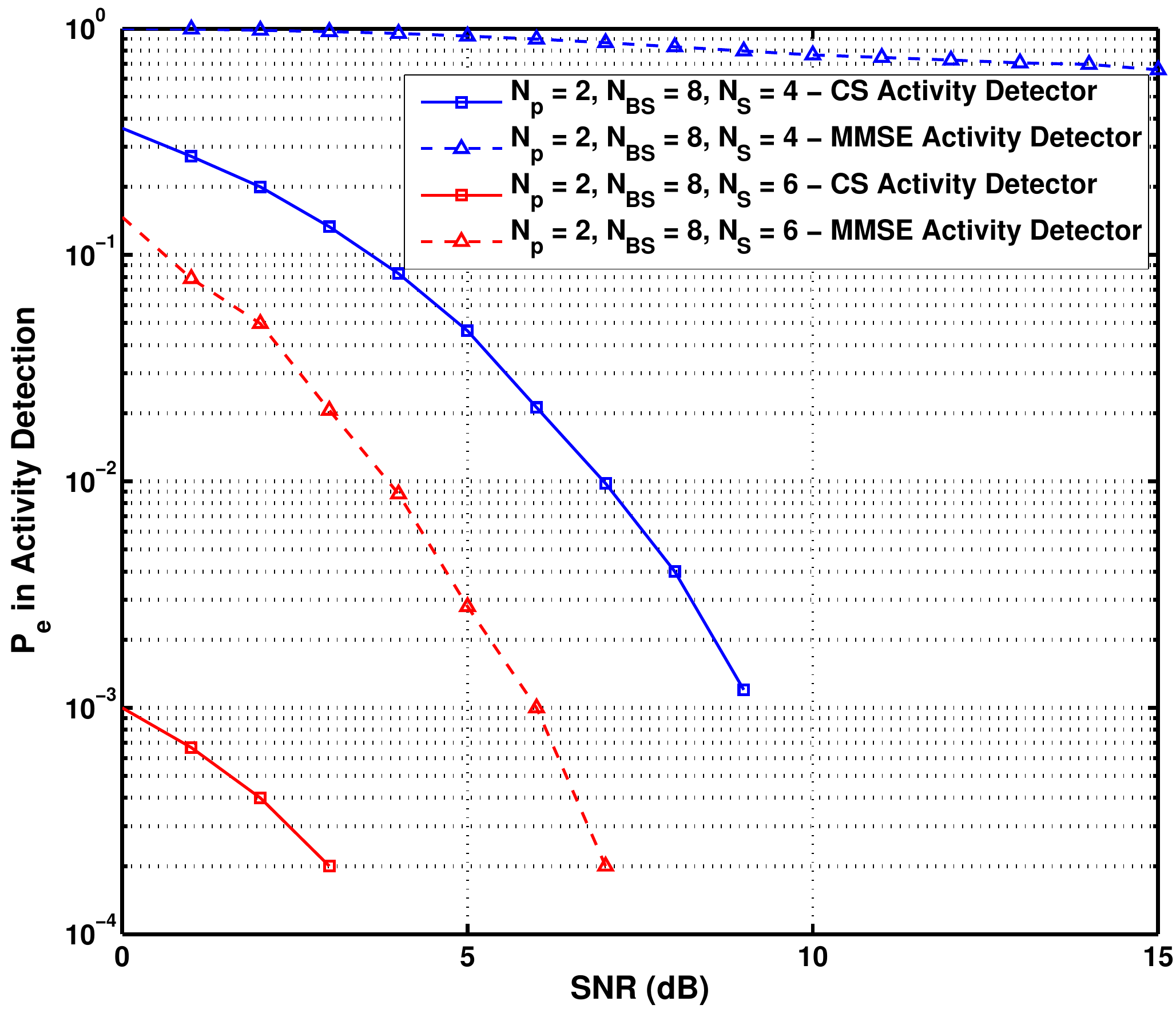}
	\caption{Activity Detection using $\ell_2/\ell_1$ CS detector and MMSE detector.}
	\label{activitydetector}
\end{figure}

\begin{figure}
	\centering
	\includegraphics[width=0.46\textwidth]{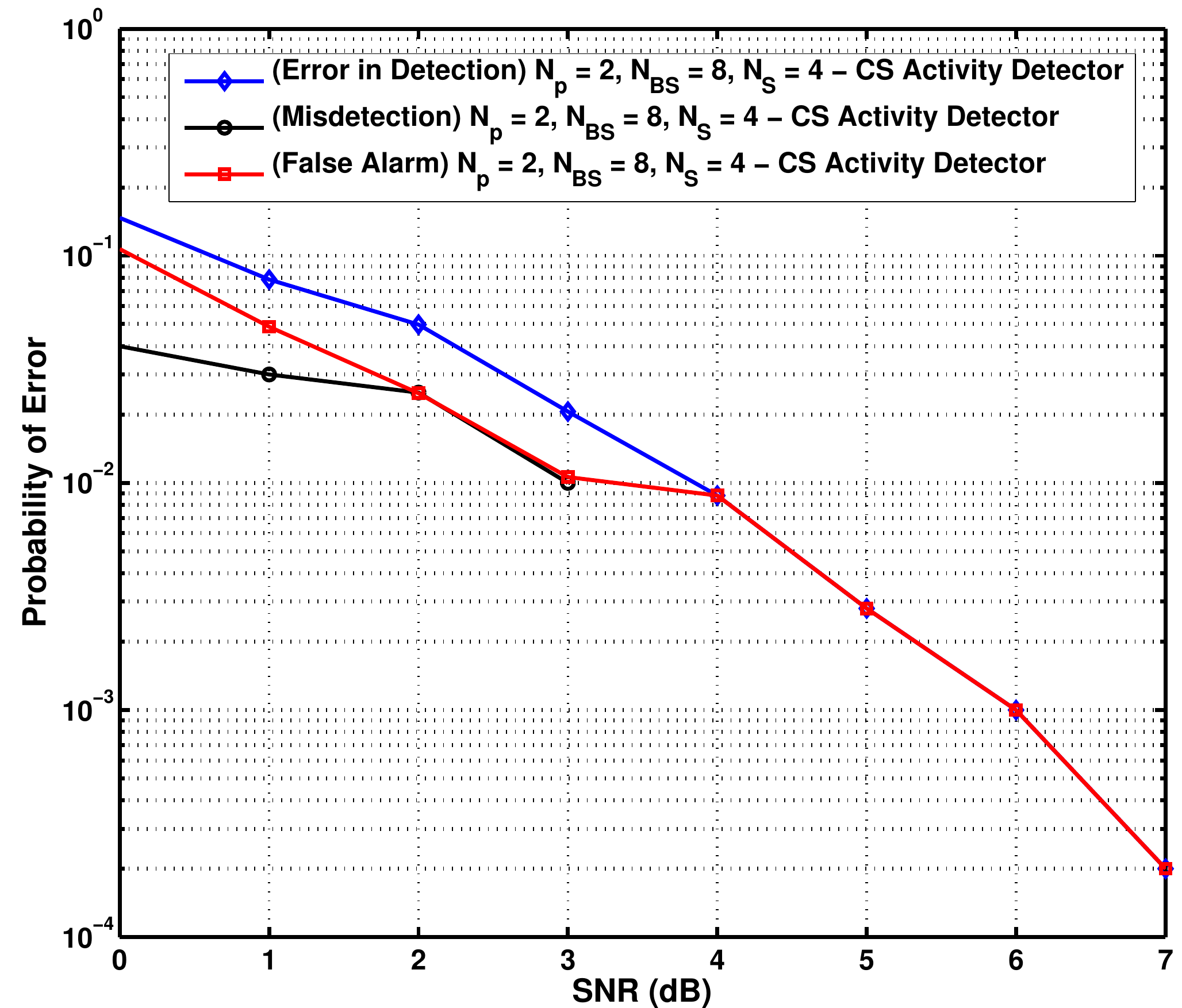}
	\caption{Probability of Error, misdetection and false alarm in
detecting Activity using CS Detector where $N_P$ = 2, $N_{BS}$ = 8, $N_S$ = 4.
}
	\label{falsealarm}
\end{figure}

Fig. \ref{activitydetector} shows the performance of the $\ell_2/\ell_1$ activity detector at different SNR conditions versus activity detected using an MMSE detector. In this simulation, we consider perfect match of the users activity states with the estimated states as no error while incorrectly detecting the activity of even a single user as an error. The simulation is repeated over 10000 iterations and the percentage of mis-detections is calculated. We used a majority rule to infer the activity of user $i$ (for both detectors), from the reconstructed signal. Every 72 elements will jointly decide the activity of a user. It can be seen from the Fig. \ref{activitydetector} that the activity detection is more reliable when using the $\ell_2/\ell_1$ detector. As the number of $N_S$ receiver antennas decrease, the performance of MMSE detector levels out while $\ell_2/\ell_1$ can still detect correct activity to a certain statistical probability of error. The figure shows the probability of error in detection for $N_S = 4,6$. An interesting observation of the CS activity detector used, is that in most erroneous cases, the detector tends to overestimate the activity of the users leading to false alarms rather than mis-detections. We coin false alarms as the detection scenarios where the detector marks $M_t$ primary users as active while the true number is $N_t$ such that $M_t$ > $N_t$. The mis-detection scenarios are situations where the number of active users are underestimated, i.e.: $M_t < N_t$. This observation is shown in Fig. \ref{falsealarm}.

\begin{figure}
	\centering
	\includegraphics[width=0.46\textwidth]{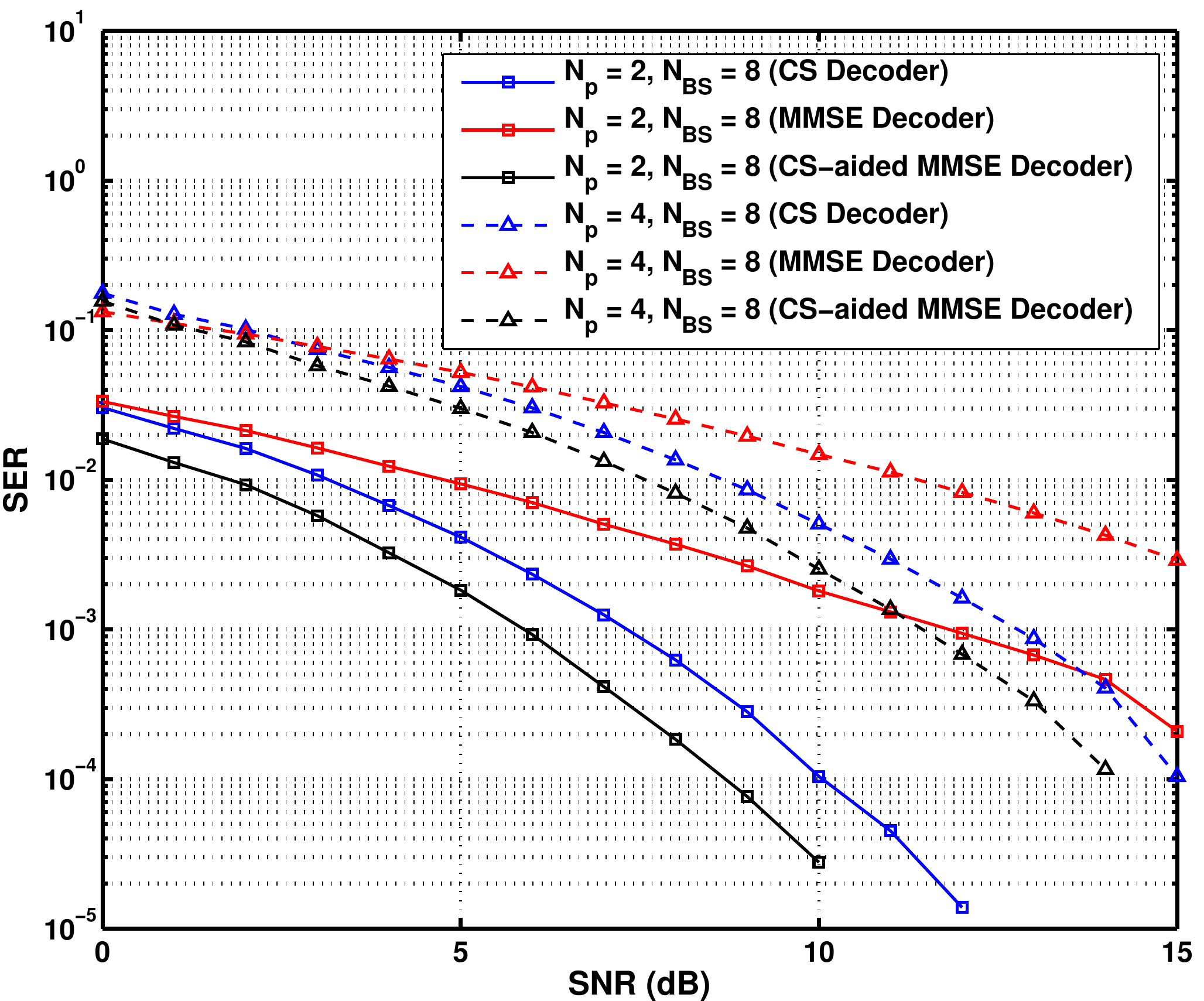}
	\caption{{\color[rgb]{0,0,0}{Symbol Error Rate (SER)}} using $\ell_2/\ell_1$ CS decoder, MMSE decoder and CS-MMSE detector for (i) 2 out of 8 active users and (ii) 4 out of 8 active users.}
	\label{ser-2-4}
\end{figure}

\begin{figure}
	\centering
	\includegraphics[height=0.38\textwidth, width=0.46\textwidth]{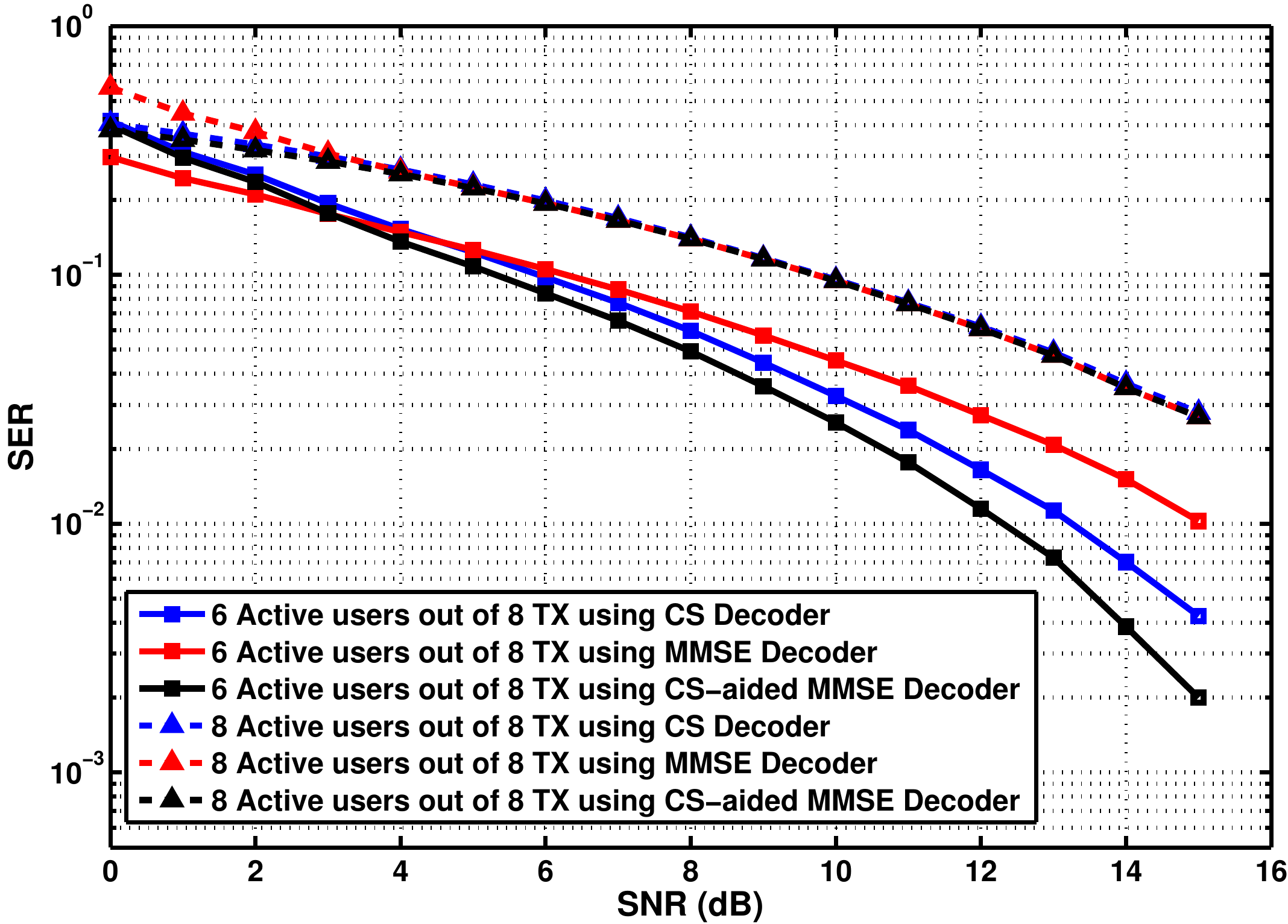}
	\caption{{\color[rgb]{0,0,0}{Symbol Error Rate (SER)}} using $\ell_2/\ell_1$ CS decoder, MMSE decoder and CS-MMSE detector for (i) 6 out of 8 active users and (ii) 8 out of 8 active users.}
	\label{ser-6-8}
\end{figure}

In Fig. \ref{ser-2-4} and Fig. \ref{ser-6-8}, we evaluate the performance of proposed CS-aided decoder by evaluating the resulting {\color[rgb]{0,0,0}{symbol error rate (SER)}} from decoding at the secondary node when the active users out of $N_P=8$ are 2, 4, 6 and 8. For comparison purposes, we include the decoding results from the stand-alone $\ell_2/\ell_1$ decoder and the MMSE decoder under the same conditions. In all cases, except from the full loaded case (8 active users), the $\ell_2/\ell_1$, having a more robust activity detection sense, yields better performance than stand-alone MMSE in low activity cases. The MMSE, $\ell_2/\ell_1$ detector and the MMSE decoder yield the same performance when the number of active users is 8. This is because when activity is detected for all users, the stand-alone MMSE problem is the same for the CS-MMSE detector. The objective function in the $\ell_2/\ell_1$ decoder is irrelevant now and the constraint is similar to the MMSE objective function, hence leads to the same conclusion.

\section{Conclusion}\label{Con}
In this paper, a generic spatial activity detector based on compressive sensing tools and reconstruction of block-sparse signals is proposed. We have shown that the proposed detector outperforms activity detection based on the MMSE estimator. Also, it has been shown that using the proposed detector to aid the MMSE estimator in a CS-MMSE model provides reliable decoding results. This reliable decoding ability can be an enabler for relaying with fewer antennas at the relay than the primary receiver(s).

\bibliographystyle{IEEEtran}
\bibliography{CSDetector}

\end{document}